\documentclass[twocolumn,showpacs,showkeys,preprintnumbers,
amsmath,amssymb,aps,floatfix,pre]{revtex4}

\usepackage{graphicx}
\usepackage{multirow}
\usepackage{times}
\usepackage{amssymb}
\usepackage{dcolumn}%
\usepackage{bm}%

\begin{document}

\title[Universal scaling]{Universal divergenceless scaling between structural relaxation and caged dynamics \\  in glass-forming systems}

\author{A.\@ Ottochian}
\affiliation{Dipartimento di Fisica ``Enrico Fermi'', 
Universit\`a di Pisa, Largo B.\@Pontecorvo 3, I-56127 Pisa, Italy}

\author{C.\@ De Michele}
\affiliation{Dipartimento di Fisica, Universit\`{a} di Roma "La Sapienza" Piazzale Aldo Moro, 2, 00185 Roma, Italy\\
and\\
INFM-CRS SOFT, Italy}

\author{D.\@ Leporini}
\email{dino.leporini@df.unipi.it}
\affiliation{Dipartimento di Fisica ``Enrico Fermi'', 
Universit\`a di Pisa, Largo B.\@Pontecorvo 3, I-56127 Pisa, Italy\\
and\\
INFM-CRS SOFT, Italy}
\date{\today}

\begin{abstract}
\noindent
On approaching the glass transition, the microscopic kinetic unit spends increasing time rattling in the cage of the first neighbours whereas its average escape time, the structural relaxation time $\tau_\alpha$, increases from a few picoseconds up to thousands of seconds. A thorough study of the correlation between $\tau_\alpha$ and the rattling amplitude, expressed by the Debye-Waller factor (DW),  was carried out. 
Molecular-dynamics (MD) simulations of both a model polymer system and a binary mixture were performed by varying the temperature, the density $\rho$, the potential and the polymer length to consider the structural relaxation as well as both the rotational and the translation diffusion. The present simulations, together with MD studies on other glassformers, evidence the scaling  between the structural relaxation and the caged dynamics. 
An analytic model of the master curve is developed in terms of two characteristic length scales $ {\overline{a^2}}^{1/2}$ and ${\sigma_{a^2}}^{1/2}$, pertaining to the distance to be covered by  the kinetic unit to reach a transition state. The model does not imply $\tau_\alpha$ divergences.
The comparison with the experiments supports the numerical evidence over a range of relaxation times as wide as about eighteen orders of magnitude. A comparison with other scaling and correlation procedures is presented. In particular, the density scaling of the length scales $ {\overline{a^2}}^{1/2}, {\sigma_{a^2}}^{1/2} \propto \rho^{-1/3}$ is shown to be not supported by the present simulations.
The study suggests that the equilibrium and the moderately supercooled states of the glassformers possess key information on the huge slowing-down of their relaxation close to the glass transition. The latter, according to the present simulations, exhibits features consistent with the Lindemann melting criterion and the free-volume model.
\end{abstract}


\maketitle

\section{Introduction}
When they are cooled or compressed, several systems like liquids, mixtures, polymers, bio-materials, metals and molten salts may avoid the crystallization and, following a huge increase of the viscosity, finally freeze into a glass, a microscopically disordered solid-like state. Understanding the extraordinary viscous slow-down that accompanies  glass formation is a major scientific challenge \cite{Angell91,Angell95,DebeStilli2001}. 

On approaching the glass transition (GT), trapping effects are more and more prominent. 
The average escape time from the cage of the first neighbors, i.e. the structural relaxation time $\tau_\alpha$, increases from a few picoseconds up to thousands of seconds. The rattling motion inside the cage  occurs on picosecond time scales with amplitude $\langle u^2\rangle^{1/2}$, the so called Debye-Waller factor (DW). The DW factor is clearly related to the short-time elastic properties of the systems \cite{Dyre06}. At first sight, due to the extreme time-scale separation between the rattling motion ($\sim 10^{-12}s$) and the relaxation ($\sim 10^2s$ at GT), one expects the complete independence of the two motions. However, already in 1943 Tobolsky, Powell, and Eyring  pointed out that there could be a relation between the curvature of the potential well near the minimum (controlling the DW factor)  and the height of the energy  barrier (limiting the flow process), thus establishing a relation between the instantaneous shear modulus $G_\infty$ and the shear viscosity $\eta$ \cite{TobolskyEtAl43}. Later, the diffusive motion was described as natural consequence of the dynamic equilibrium between vibrational and configurational quantum states \cite{Angell68} and the free-energy barrier for viscous flow was found as being proportional to $G_\infty(T)$ \cite{Nemilov68}. 

A firmer basis to connect fast and slow degrees of freedom was developed by Hall and Wolynes who, assuming that atomic motion is restricted to cells, pictured the glass transition as a freezing in an aperiodic crystal structure (ACS) modeled by the density functional theory \cite{HallWoly87}.
As a result, the viscous flow is described in  terms of  activated jumps over energy barriers $\Delta E \propto k_B T a^2/ \langle u^2\rangle$ where $a$ is the displacement to reach the transition state and $k_B$ the Boltzmann constant. The usual rate theory leads to the Hall-Wolynes equation (HW):
\begin{equation}
\label{dyreWolynes}
\tau_\alpha^{(HW)}, \eta^{(HW)} \propto exp\left(\frac{a^2}{2\langle u^2\rangle}\right)
\end{equation}
The ACS model is expected to fail when $\tau_{\alpha}$ becomes comparable to the typical rattling times of each atom in the cage of the surrounding atoms, corresponding to picosecond timescales. That condition is quite mild, e.g. in Selenium it occurs at $T_m + 104 K$ ($T_m$ is the melting temperature) \cite{BuchZorn92}. Buchenau and Zorn derived a relation very similar to Eq.\ref{dyreWolynes} in terms of soft vibrational modes \cite{BuchZorn92}:
\begin{equation}
\label{BuchZorn}
\tau_\alpha, \eta \propto \exp \left (\frac{u_0^2}{2\langle u^2\rangle_{loc}}\right )
\end{equation}
where $u_0$ is a critical displacement to allow for the elementary flow or $\alpha$-relaxation process and $\langle u^2\rangle_{loc}$ is the difference between the DW factor in the liquid phase $\langle u^2\rangle$ and its extrapolation from the low-temperature values. The definition of $\langle u^2\rangle_{loc}$ affects the plot $\log\ \eta$ vs. $1/\langle u^2\rangle_{loc}$. If the extrapolation of either the glass or the crystal contribution is subtracted from the DW factor of selenium, a convex curve or a straight line are seen, respectively \cite{BuchZorn92}. The fact that many glass-formers have no underlying crystalline phases, as well as the fact that in other studies removing the glass contribution, differently from selenium, the plot $\log\,\eta$ vs. $1/\langle u^2\rangle_{loc}$ is a straight line \cite{Paciaroni05,Magazu04}, raises some ambiguities about the above subtractions.
Buchenau and Zorn also noted  that, if no subtraction is made, the curve  $\log\ \eta$ vs. $1/\langle u^2\rangle$ for selenium is concave, namely the HW equation, Eq.\ref{dyreWolynes}, is not obeyed. The HW equation has been derived in the framework of the so called elastic models (for a review see ref.\cite{Dyre06}), like the shoving model \cite{Dyre04,Dyre96}. 
 
 The HW equation states that the glass softens when the DW factor exceeds a critical value, which is reminiscent of the Lindemann melting criterion for crystalline solids \cite{Bilgram87}, The empirical law $T_g \simeq 2/3 T_m$ ($T_g$ is the glass-transition  temperature) \cite{Angell91, DebeStilli2001,Ngai00} also suggests that the melting and the glass transition have a common basis. This viewpoint led to an alternative derivation of Eq.\ref{dyreWolynes} \cite{NoviSoko03} and motivated extensions of the Lindemann criterion to glasses \cite{XiaWolynes00}. The closeness of the HW equation with free-volume concepts \cite{Gedde95} was noted \cite{HallWoly87} and investigated numerically \cite{StarrEtAl02}. 
 
Other studies noted a relation between the fast vibrational dynamics and the long-time relaxation both far \cite{Harrowell06,Sastry01} and close to the glass transition \cite{MarAngell01,Angell95B,Angell95,Ngai04,Ngai00}. A numerical investigations pointed out that the short-time DW heterogeneities predict the spatial distribution of the long-time dynamic propensities \cite{Harrowell06}. The fragility $m$, a steepness index of how fast the viscosity $\eta$ or $\tau_\alpha$ increase {\it close} to $T_g$ \cite{Angell91}, has been also considered. It has been proposed that variations in the fragility  originate in differences between their vibrational heat capacities, harmonic and anharmonic \cite{MarAngell01} and depend on changes in the vibrational properties of individual energy minima of the energy landscape in addition to their total number and spread in energy \cite{Sastry01}. The temperature dependence of the DW factor around the glass transition has been also studied \cite{Angell95B,Angell95,Ngai04,Ngai00}. It was seen that for strong glassformers (small fragility) DW is almost linear with temperature, whereas  a stronger than linear dependence takes place for fragile systems pointing to increasing anharmonicity of the short-time dynamics. With a distinct approach further studies established correlations between the vibrational dynamics and the relaxation {\it close} to the glass transition, as quantified by the fragility \cite{ScopignoEtAl03,SokolPRL,Buchenau04,NoviSoko04,NovikovEtAl05} with controversies \cite{Johari06}. Finally, as further examples of studies comparing the fast and the slow dynamics, we point out the correlations between the structural and the secondary relaxations in a supercooled liquid \cite{Bohmer06}, as well as between the apparent activation energy above the glass transition and the fragility \cite{NoviSoko04,NovikovEtAl05}. 
 
In a recent paper we reported the universal dependence between the structural relaxation time and the DW factor for a model polymer \cite{OurNatPhys}. The universal scaling curve, which is described by a simple generalization of the HW equation (Eq.\ref{dyreWolynes}), fits with the existing experimental data from supercooled liquids, polymers and metallic glasses over about eighteen decades of relaxation times and a very wide range of fragilities. Here we show by novel numerical simulations that the scaling holds for binary mixtures with different interacting potentials, density and temperatures, i.e. for an {\it atomic, heterogeneous} system different by the molecular, homogeneous one considered in ref.\cite{OurNatPhys}. Moreover, we prove that it holds not only for the translational degrees of freedom but for the rotational ones of the model polymer as well. Comparisons with other  numerical studies \cite{MSD_SiO,ISF_SiO,IcosahedralGlotzer, NanoparticleDouglas, OTPRuocco}, novel  experimental data \cite{GeO2Fontana} as well as other scaling and correlation procedures is also presented.
 
The paper is organized as follows: in Sec.\ref{genHW} the HW equation is suitably generalized.  In Sec.\ref{methods} the numerical methods  are described. The results are presented and discussed in Sec.\ref{ResDis}. The conclusions are given in Sec.\ref{Conclusions}.

\section{Generalized Hall-Wolynes equation}
\label{genHW}
\noindent
One basic assumption of  the original HW equation, Eq.\ref{dyreWolynes}, is that the distance to reach the transition state  has a characteristic value $a$. Actually, this length scale is dispersed. To model the related distribution, it is assumed that the latter does not depend on the state parameters such as the temperature, the density or the interacting potential. This complies with the spirit of ref.\cite{HallWoly87} where the  $a$ distance is said to be mostly controlled by the geometrical packings. It is also known  that, irrespective of the relaxation 
time $\tau_\alpha$, the average distance moved by the relaxing unit within $\tau_\alpha$ is about the
same, i.e. a fraction of the molecular diameter \cite{Angell91}. As a suitable choice, the distribution of the squared distances $p(a^2)$ is taken as  a truncated gaussian form 
\begin{equation}
p(a^2) = \left\{ \begin{array}{ll}
A \exp\left (-\frac{(a^2 - \overline{a^2})^2}{2\sigma^2_{a^2}}\right )
 & \textrm{if $a>a_{min}$}\\
0& \textrm{otherwise}
\end{array} \right.
\label{Eq:gaussianDistro}
\end{equation}
where $A$ is the normalization and $a^2_{min}$ is the minimum displacement to reach the transition state. Averaging the HW Eq.\ref{dyreWolynes} over the distribution given by Eq.\ref{Eq:gaussianDistro}, yields the following generalized HW equation (GHW):
\begin{eqnarray}
\tau_\alpha &=& \int_0^\infty d a^2 p(a^2) \, \tau_\alpha^{(HW)}(a^2)  \label{Eq:GHWtrunc0}\\
&=& B \; {\cal N}[\langle u^2 \rangle]  \exp\left( \frac{\overline{a^2}}{2\langle u^2\rangle } + \frac{ \sigma^2_{a^2}}{8\langle u^2\rangle ^2 } \right )
\label{Eq:GHWtrunc1}
\end{eqnarray}
where $B$ is a constant and the normalization factor ${\cal N}[\langle u^2 \rangle]$ reads:
\begin{equation}
{\cal N}[\langle u^2 \rangle] = \frac{1 + \hbox{Erf} \left [  \frac{( \overline{a^2} - a^2_{min})/\sigma_{a^2}+ \sigma_{a^2}/2\langle u^2\rangle}{\sqrt{2}} \right ]}{1 + \hbox{Erf} \left [ \frac{(\overline{a^2} - a^2_{min})/\sigma_{a^2}}{\sqrt{2}}\right ] } 
\label{Eq:ErfFact}
\end{equation}
If $\overline{a^2} \ge a^2_{min}$,  $ 1 \le {\cal N}[\langle u^2 \rangle] \le 2$, namely ${\cal N}[\langle u^2 \rangle]$ depends very weakly on the DW factor $\langle u^2\rangle$ and the influence of the truncation is negligible. Then, $\tau_0 \equiv B {\cal N}[\langle u^2 \rangle] \simeq const $ and Eq.\ref{Eq:GHWtrunc1} reduces to:
\begin{equation}
\tau_\alpha =  \tau_0 \;  \exp\left( \frac{\overline{a^2}}{2\langle u^2\rangle } + \frac{ \sigma^2_{a^2}}{8\langle u^2\rangle ^2 } \right )
\label{parabola}
\end{equation}
An analogous law holds for the viscosity $\eta$. Owing to the finite value of the DW factor, Eq.\ref{parabola} does not imply the divergence of $\tau_\alpha$ \cite{DyreNatPhys08,McKenna08}.

The motivations behind the gaussian form of $p(a^2)$ mainly rely on the Central Limit Theorem. In fact, $a^2$ ($r^2_0$ in the notation of ref.\cite{HallWoly87}) is the cumulative displacement of the $N_m$ particle that move \cite{HallWoly87}. Other supporting facts for the gaussian form of $p(a^2)$ are the following. If the kinetic unit performs harmonic oscillations around the equilibrium position with an effective spring constant $k$, the DW factor becomes $\langle u^2\rangle = k_B T/ k $ and Eq. (\ref{parabola}) reduces to: 
\begin{equation}
\tau_\alpha = \tau_0\exp\left(\frac{k \overline{a^2}}{2 k_B T} + \frac{k^2 \sigma^2_{a^2}}
{8 (k_B T)^2} \right)
\label{Eq:harm}
\end{equation}
The above expression  was reported for both supercooled liquids \cite{Bassler87} and 
polymers \cite{FerryEtAl53}. Along a similar line of reasoning, assuming harmonic oscillations leads to the following expression for the energy barrier height $\Delta E$ \cite{Dyre06}:
\begin{equation}
\Delta E = \frac{1}{2} k a^2
\label{Eq:harmDelE}
\end{equation} 
Eq.\ref{Eq:harmDelE} allows one to reinterpret the gaussian form of $p(a^2)$ as a gaussian distribution of energy barriers \cite{MonthBouch96}.
Substituting Eq. (\ref{Eq:harmDelE}) into Eq. (\ref{Eq:harm}), one recovers a key result of the facilitated model of glass-formers developed by Garrahan and Chandler \cite{EastArrow}:
\begin{equation}
\tau_\alpha = \tau_0\exp\left(\frac{\overline{\Delta E}}{k_B T} + \frac{ \sigma^2_{\Delta E}}{2 (k_B T)^2} \right)
\label{Eq:GaussBarrHW}
\end{equation}

\section{Methods}
\label{methods} 
\subsection{Models}
\subsubsection{Polymer melt}
A coarse-grained model of a linear polymer chain is used. Torsional potentials
are neglected. We considered a system of $N_{m}=2000$ monomers in all cases but $M=3$ where $N_{m}=2001$. 
Non-bonded monomers at a distance $r$ interact via the truncated parametric potential:
\begin{equation}
U_{q,p}(r) = \frac{\epsilon}{p-q} \left [  p \left (\frac{\sigma^*}{r}\right )^{q} - q \left (\frac{\sigma^*}{r}\right )^{p}\right ] + U_{cut}
\label{Eq:modifiedLJ}
\end{equation}
where $\sigma^* = 2^{1/6} \sigma$ and the value of the constant $U_{cut}$ is chosen to ensure $U_{p,q}(r) = 0$ at $r \ge r_c = 2.5 \sigma $. The minimum of the potential $U_{p,q}(r)$ is at 
$r = \sigma^*$, with a constant depth $U(r = \sigma^*) = \epsilon$. Note that $U_{q,p}(r) = U_{p,q}(r)$. 
Bonded monomers interact with a potential which is the sum of the FENE (Finitely Extendible Nonlinear Elastic) potential and the Lennard-Jones (LJ) potential \cite{sim}. The resulting bond length is $b = 0.97\sigma$ within few percent.
We set $\sigma = 1$, $\epsilon = 1$. The time unit is $\tau_{MD} = (m\sigma^2 / \epsilon)^{1/2}$, with $m$ being the mass of the monomer. 
Temperature is in units of $\epsilon/k_B$, where $k_B$ is the Boltzmann constant. We set $m = k_B = 1$.
NPT and NTV ensembles have been used for equilibration runs while NVE ensemble has been used for production runs  for a given state point  (labelled by the multiplets $\{T,\rho,M,p,q\}$).
NPT and NTV ensembles have been simulated with the extended system method introduced by Andersen \cite{NPTanders} and Nos\'e \cite{NTVnose}. The numerical integration of the augmented Hamiltonian has been performed through the reversible  
multiple time steps algorithm (r-RESPA algorithm) {\it et al.}\cite{respa}.
In particular, the NPT and NTV Liouville operators have been factorized using the Trotter theorem \cite{trotter} separating the short range and 
long range contributions of the potential $U_{p,q}(r)$ (see Eq. \ref{Eq:modifiedLJ}), according to the WCA decomposition \cite{tuckWCA}.

\subsubsection{Binary mixtures}
An 80:20 binary mixture (BM) of $N_{bm}=1000$ particles is considered. The two species are labelled $A$, $B$ and particles interact via the potential:
\begin{equation}
U_{q,p,\alpha,\beta}(r) = \frac{\epsilon_{\alpha,\beta}}{p-q} \left [  p \left (\frac{\sigma_{\alpha,\beta}^*}{r}\right )^{q} - q \left (\frac{\sigma_{\alpha,\beta}^*}{r}\right )^{p}\right ] + U_{cut}
\label{Eq:modifiedLJmix}
\end{equation}
that is similar to Eq. (\ref{Eq:modifiedLJ}), except that the well height 
and the minimum of the potential now depend on the interacting species, being
$\alpha,\beta \in A,B$ with $\sigma_{AA}=1.0$, $\sigma_{AB}=0.8$, 
$\sigma_{BB}=0.88$, $\epsilon_{AA}=1.0$, $\epsilon_{AB}=1.5$, $\epsilon_{BB}=0.5$.
Note that setting $q=12$, $p=6$ in Eq. (\ref{Eq:modifiedLJmix}) and the above choices for $\sigma_{\alpha,\beta}$ and $\epsilon_{\alpha,\beta}$,
reduce the model  to the well-known LJ Kob-Andersen model (BMLJ)\cite{KobAndersenPRE1,KobAndersenPRE2,KobAndersenPRL}. 
The system was equilibrated in the NTV ensemble and the production runs were carried out in the NVE ensemble.  NTV runs used a standard Nos\'e method 
\cite{NTVnose}. The ''velocity verlet'' integration algorithm was used both in the NVE and NVT ensembles \cite{allentildesley}.

\section{Results and discussion}
\label{ResDis}

\subsection{Relaxation and transport properties}
\label{relax}
First, the monomer dynamics has been studied. To this aim, one defines the mean square displacement (MSD) $\langle r^2(t)\rangle$ as:
\begin{equation}
\langle r^2(t)\rangle = \frac{1}{N} \sum_i \langle \|{\bf x}_i(t) - {\bf x}_i(0)\|^2  \rangle
\label{Eq:MSD}  
\end{equation}
In addition to MSD the self part of the intermediate scattering function (ISF) is also considered:
\begin{equation}
F_{s}(q,t) = \frac{1}{N} \langle \sum_j^N e^{i{\bf q}\cdot({\bf x}_j(t) - {\bf x}_j(0))} 
\label{Eq:Fself}
\rangle 
\end{equation}
ISF was evaluated at $q= q_{max}$, the maximum of the static structure factor. $N=N_m$ and $N=N_{bm}$ for polymers and binary mixtures, respectively. ${\bf x}_i$ is the position of the $i$-th monomer (polymers) or particle (BM). Note that for BM MSD is averaged over both species $A$ and $B$. 

\begin{figure}[t]
\begin{center}
\includegraphics[width=8cm]{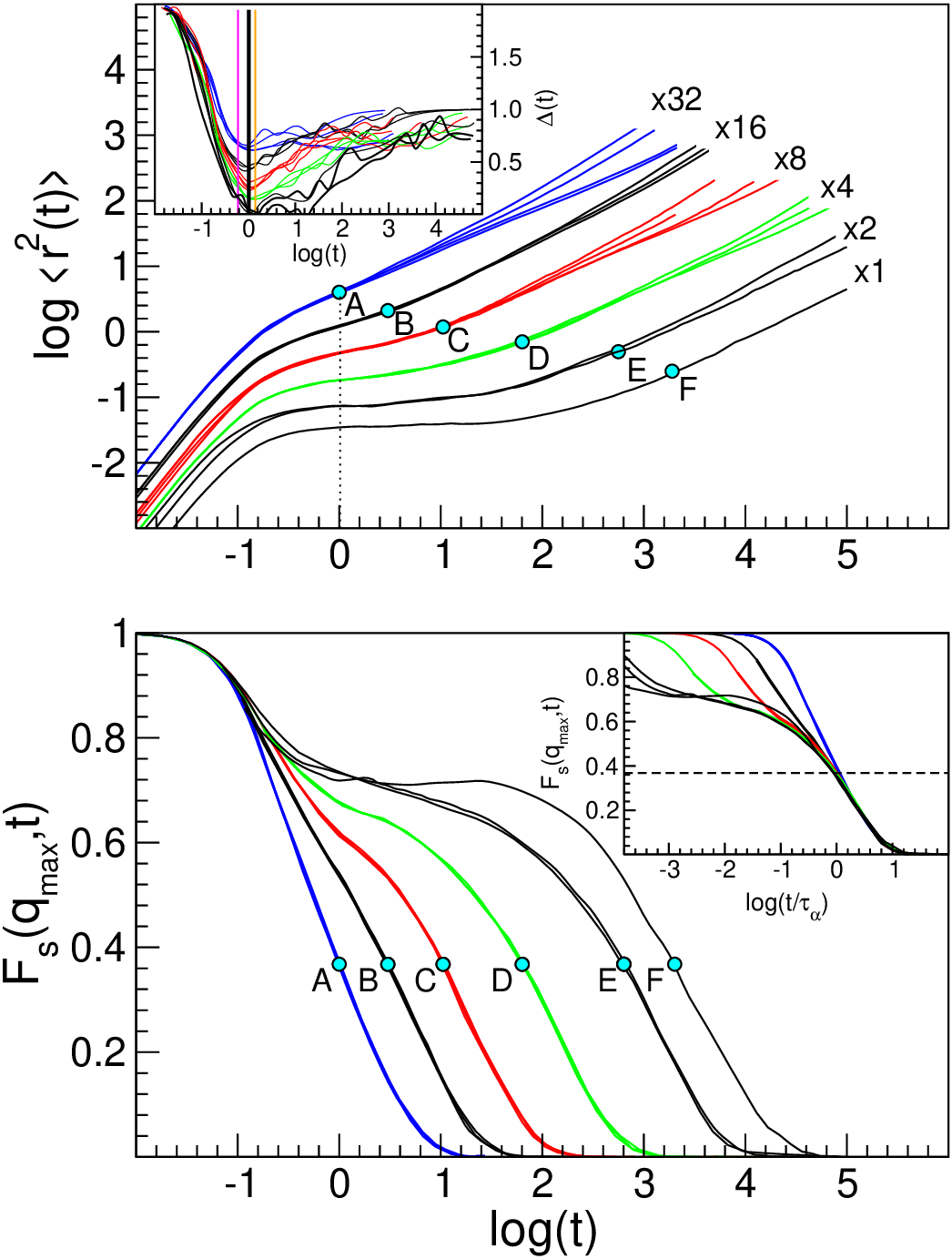}
\end{center}
\caption{Monomer dynamics in the polymer melt. Top: MSD time-dependence for polymers in selected cases. MSDs are multiplied by indicated factors. Inset: corresponding MSD slope $\Delta(t)$; the uncertainty range on the position of the minimum at $t^{\star} = 1.0(4)$ (black line) is bounded by the vertical colored lines.  Bottom: corresponding ISF curves
for polymers. Inset: superposition of the ISF curves. Four sets of clustered curves (A through D) show that, if states have equal $\tau_\alpha$ (marked with dots on each curve), the MSD and ISF curves coincide from times fairly longer than $\tau_{\alpha}$ down to the crossover to the ballistic regime at least.}
 \label{fig1}
\end{figure}

Fig.\ref{fig1} shows typical MSD and ISF curves of the polymeric monomers. At very short times (ballistic regime) MSD increases according to $\langle r^2(t)\rangle \cong (3 k_B T/m) t^2$ and ISF starts to decay. The repeated collisions with the other monomers slow the displacement of the tagged one, as evinced by the knee of MSD at $ t \sim \sqrt{12}/\Omega_0 \sim 0.17$, where $\Omega_0$ is an effective collision frequency, i.e. it is  the mean small-oscillation frequency of the monomer in the potential well produced by the surrounding ones kept at their equilibrium positions \cite{Boon}.
At later times a quasi-plateau region, also found in ISF, occurs when the temperature is lowered and/or the density increased. This signals the increased caging of the particle. The latter is released after an average time $\tau_{\alpha}$, defined by the relation $F_s(q_{max}, \tau_{\alpha}) = e^{-1}$. For $t \gtrsim \tau_{\alpha}$ MSD increases more steeply. The monomers of short chains ($M \lesssim 3$) undergo diffusive motion $\langle r^2(t)\rangle \propto t^\delta$ with $\delta=1$. For longer chains,  owing to the increased connectivity,  the onset of the diffusion is preceded by a subdiffusive region ($\delta < 1$, Rouse regime) \cite{Gedde95}. 

\begin{figure}[t]
\begin{center}
\includegraphics[width=8cm]{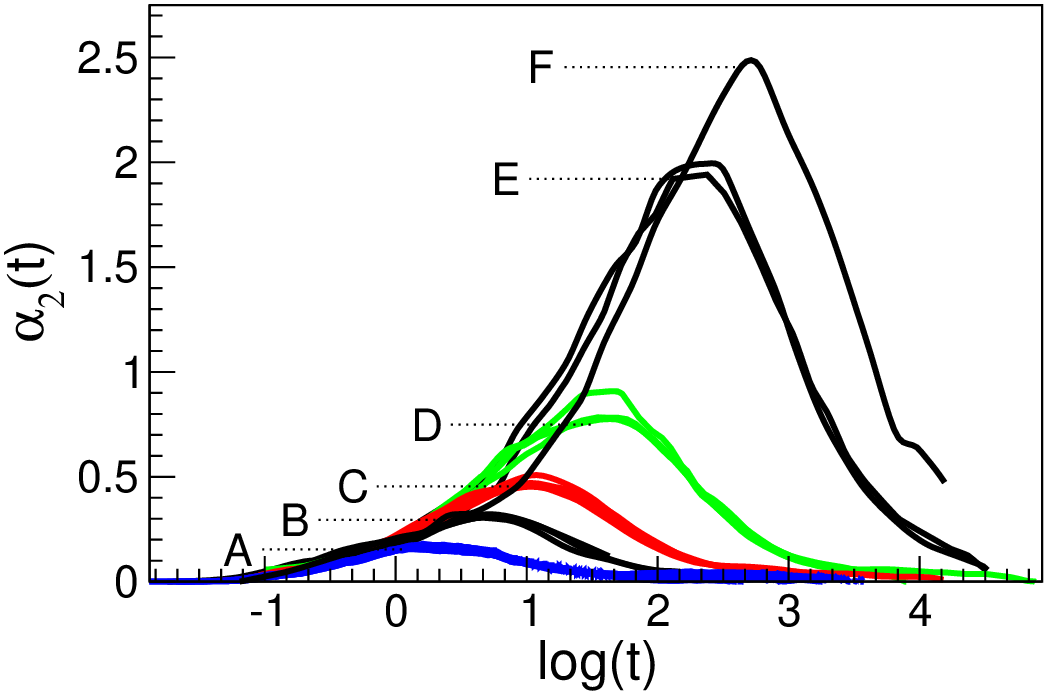}
\end{center}
\caption{\label{fig2}The non-gaussian parameter of the polymer states plotted in Fig.\ref{fig1}. }
\end{figure}

The monomer dynamics depends in a complex way on the state parameters. 
Nonetheless, if two states (labelled by multiplets $\{ T, \rho, M, p,q \}$) have equal relaxation time $\tau_{\alpha}$, the corresponding MSD and ISF curves coincide from times fairly longer than $\tau_{\alpha}$ down to the crossover to the ballistic regime and even at shorter times if the states have equal temperatures. Examples are shown in Fig.\ref{fig1}. See EPAPS supplementary material at {\bf [URL will be inserted by AIP]} for the details on all the investigated states.
Notice that the coincidence of MSD and ISF curves of states with equal $\tau_{\alpha}$ at intermediate times ($t \lesssim \tau_{\alpha}$) must not be confused with the customary superposition of ISF curves at long times ($t \gtrsim  \tau_{\alpha}$) following a suitable logaritmic time shift (see the lower-panel inset of Fig.\ref{fig1}). States with coinciding MSD and ISF have close non-gaussian properties \cite{NGP_Wolynes}. This is shown by the non-gaussian parameter (NGP):
\begin{equation}
\alpha_2(t) =  \frac{3}{5} \frac{\langle r^4(t)\rangle}{\langle r^2(t)\rangle^2} - 1 
\label{Eq:alpha2}
\end{equation}
where $\langle r^4(t) \rangle$ is defined analogously to MSD, Eq.\ref{Eq:MSD}. Plots of $\alpha_2(t)$ for the same states of Fig.\ref{fig1} are shown in Fig.\ref{fig2}. One notices that states with coinciding MSD and ISF have coinciding NGP as well. 

\begin{figure}[t]
\begin{center}
\includegraphics[width=8cm]{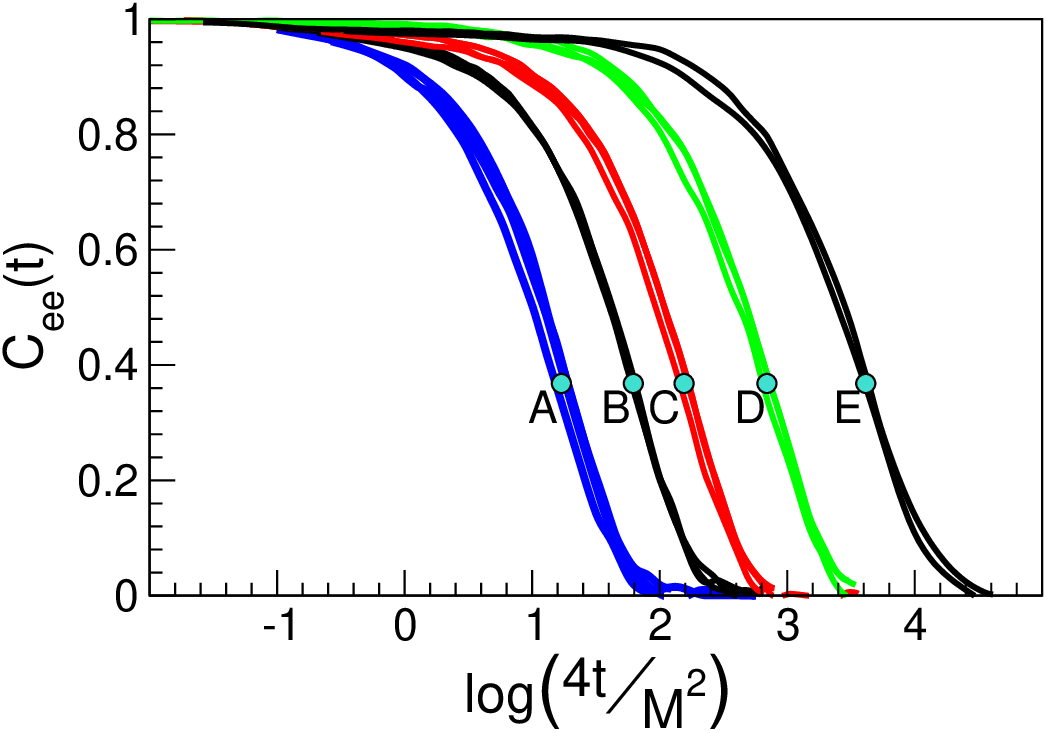}
\end{center}
\caption{Correlation functions of the end-to-end vector of the states of Fig.\ref{fig1}. The scaled time removes the chain length dependence. Polymer states contributing to one cluster of scaled curves have not necessarily equal chain length. Dots mark the time $4 \tau_{ee}/ M^2$.}
\label{fig3}
\end{figure}

Owing to the fully-flexible character of the chain, the structural relaxation time little depends on the chain length $M$ \cite{BarbieriEtAl2004}. Much stronger dependence is expected for the diffusion coefficient and the reorientation time of the whole chain which for unentangled chains ($M \lesssim 32$ {   }\cite{sim}) scale as  $ D^{-1} \propto M$ and  $\tau_{ee} \propto M^2$, respectively \cite{DoiEdwards}. These processes set the long-time dynamics of the chain and it is interesting to see if states with coinciding MSD, ISF and NGP, involving short and intermediate time scales, also exhibit coinciding translational and rotational diffusion. To this aim, the global rotational dynamics of the chain has been investigated by the correlation function of the end-to-end vector:
\begin{equation}
C_{ee}(t) = \frac{1}{N_m R_{ee}^2} \sum_{i=1}^{N_m} \langle {\bf R}_i(t)\cdot {\bf R}_i(0) \rangle
\end{equation}
where ${\bf R}_i(t)$ is the vector joining the first and the last monomer (i.e. the end monomers) of the $i$-th polymer in the melt and 
\begin{equation}
R_{ee}^2 = \frac{1}{N_m} \sum_{i=1}^{N_m} \|{\bf R}_i\|^2 
\label{Eq:meanSqEEdist}
\end{equation} 
$C_{ee}(t)$ monitors the {\it collective} relaxation, whereas ISF the single-particle one. One defines $\tau_{ee}$ via the equation $C_{ee}(\tau_{ee})=1$.  For unentangled polymers $\tau_{ee} \sim 4 M^2 \tau_{\alpha}$ \cite{DoiEdwards}. Fig.\ref{fig3} plots the correlation function $C_{ee}$ in dependence of the reduced time $4 t/ M^2$, i.e. the curves are scaled onto the one of the dimer ($M=2$) whose  end-to-end vector is the bond itself. Having removed the chain-length dependence by proper rescaling, the states with coinciding MSD, ISF and NGP exhibit coinciding end-to-end correlation loss $C_{ee}(t)$ too.  Note that the polymer states contributing to one cluster of scaled curves have not  necessarily equal chain length (see EPAPS supplementary material at {\bf [URL will be inserted by AIP]} for the details on all the investigated states). 
This is also evidenced by inspecting Fig.\ref{fig1}(top). In fact, up to $t \sim \tau_{\alpha}$ the connectivity effects are negligible and, irrespective of the $M$ value, MSD curves coincide, whereas at longer times monomer bonding comes into play and the curves start to differ from each other due to the different chain lengths \cite{DoiEdwards}.  It must be noted that, since $D  \propto R_{ee}^2/ \tau_{ee}$ \cite{CapacciEtAl04}, the collapse of the correlation function $C_{ee}$ ensures that the quantity $D \cdot M$ is identical for states with coinciding MSD, ISF and NGP.

\begin{figure}[t]
\begin{center}
\includegraphics[width=8cm]{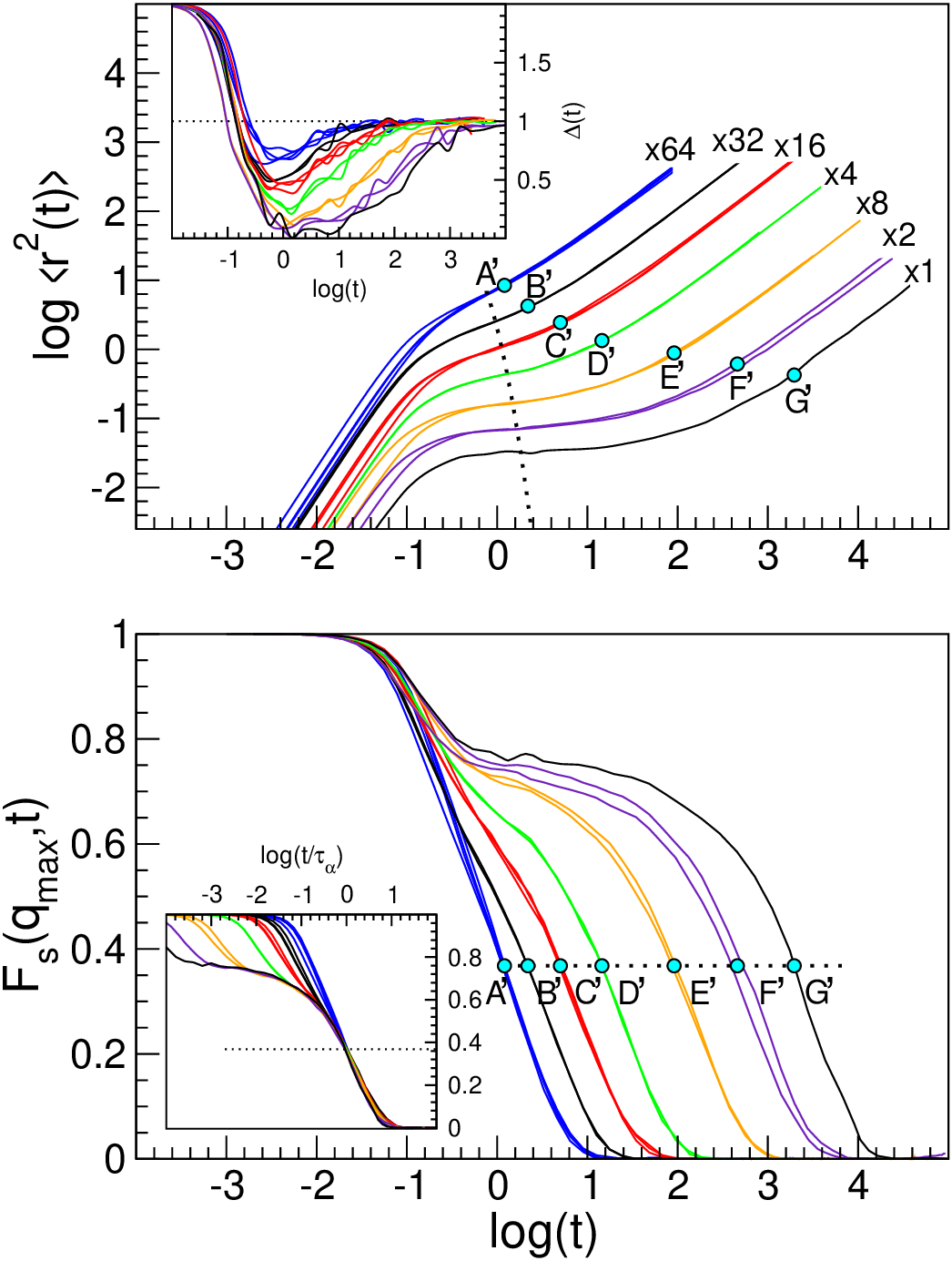}
\end{center}
\caption{ Average particle dynamics in the BM system. Top: MSD of selected cases. MSDs are multiplied by the indicated factors.  The dashed line shows the position of the minimum of $\Delta(t)$, which is plotted  in the inset. Note that it shifts at longer times for states with slower relaxation.
Bottom: corresponding ISF curves. Inset: superposition of the ISF curves. Four sets of clustered curves (A through D) show that, if states have equal $\tau_\alpha$ (marked with dots on each curve), MSD and ISF curves coincide at least from the end of the ballistic regime onwards. If the temperatures are the same, the coincidence include the ballistic regime too.}
\label{fig4}
\end{figure}

We now consider the BM system. Fig.\ref{fig4} shows typical MSD and ISF curves. Note that for the BM system these quantities are averaged over both $A$ and $B$ species.  One sees that, if two states (labelled by multiplets $\{ T, \rho, p,q \}$) have equal relaxation time $\tau_{\alpha}$, the corresponding MSD and ISF curves coincide at least from the end of the ballistic regime onwards, i.e. the states have {\it equal} diffusion coefficients $D = \lim_{t \to \infty} \langle r^2(t)\rangle/6$. This shows that, due to the missing connectivity of BM,  the MSD coincidence is not interrupted at $t \gtrsim \tau_{\alpha}$ as it happens in polymers (see Fig.\ref{fig1}). Remarkably, if the temperatures of the BM states are the same the MSD and ISF coincidence include the ballistic regime too.
\begin{figure}[t]
\begin{center}
\includegraphics[width=8cm]{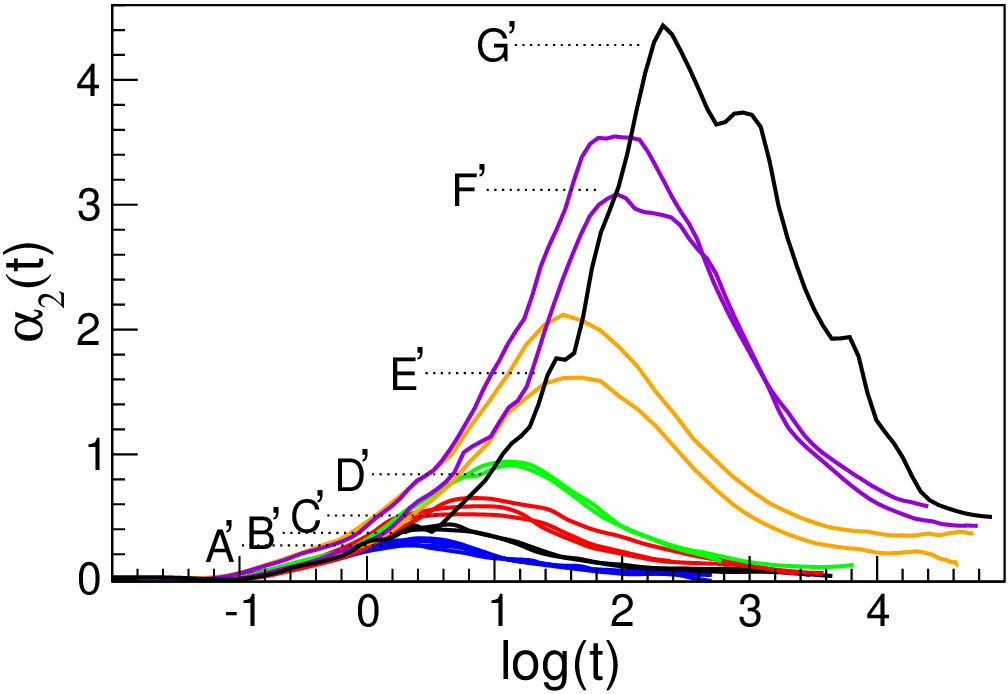}
\end{center}
\caption{ The non-gaussian parameter of the BM states of Fig.\ref{fig4}.}
\label{fig5}
\end{figure}
In full analogy with the polymer case Fig.\ref{fig5} shows that states with coinciding MSD and ISF have coinciding NGP too. 

\subsection{Scaling between relaxation and caged dynamics}
\label{scaling}

\subsubsection{Polymers and Binary Mixtures}
\label{scalingPolBM}

The results of Sec.\ref{relax} strongly support the conclusion of a close correlation between the caged dynamics at short times and the long-time dynamics, including both the structural relaxation, the chain reorientation and the diffusivity. In order to better evidence such a correlation a suitable metrics of the caged dynamics is needed. This is achieved by considering the Debye-Waller (DW) factor  $\langle u^2\rangle$, a characteristic length scale of the particle temporarily trapped into the cage.

Preliminarily, one has to clarify if the cage exists. From this respect, it must be pointed out that  the product  $\Omega_0 \tau_{\alpha}$ is  $\sim 20 $ for states with the fastest structural relaxation, meaning that the structural relaxation time is at least one order of magnitude longer than the collision time. Furthermore, in the present study  the time velocity correlation function (VCF), after a first large drop due to pair collisions, reverses the sign since the monomer rebounds from the cage wall (data not shown).

\begin{figure}[t]
\begin{center}
\includegraphics[width=8cm]{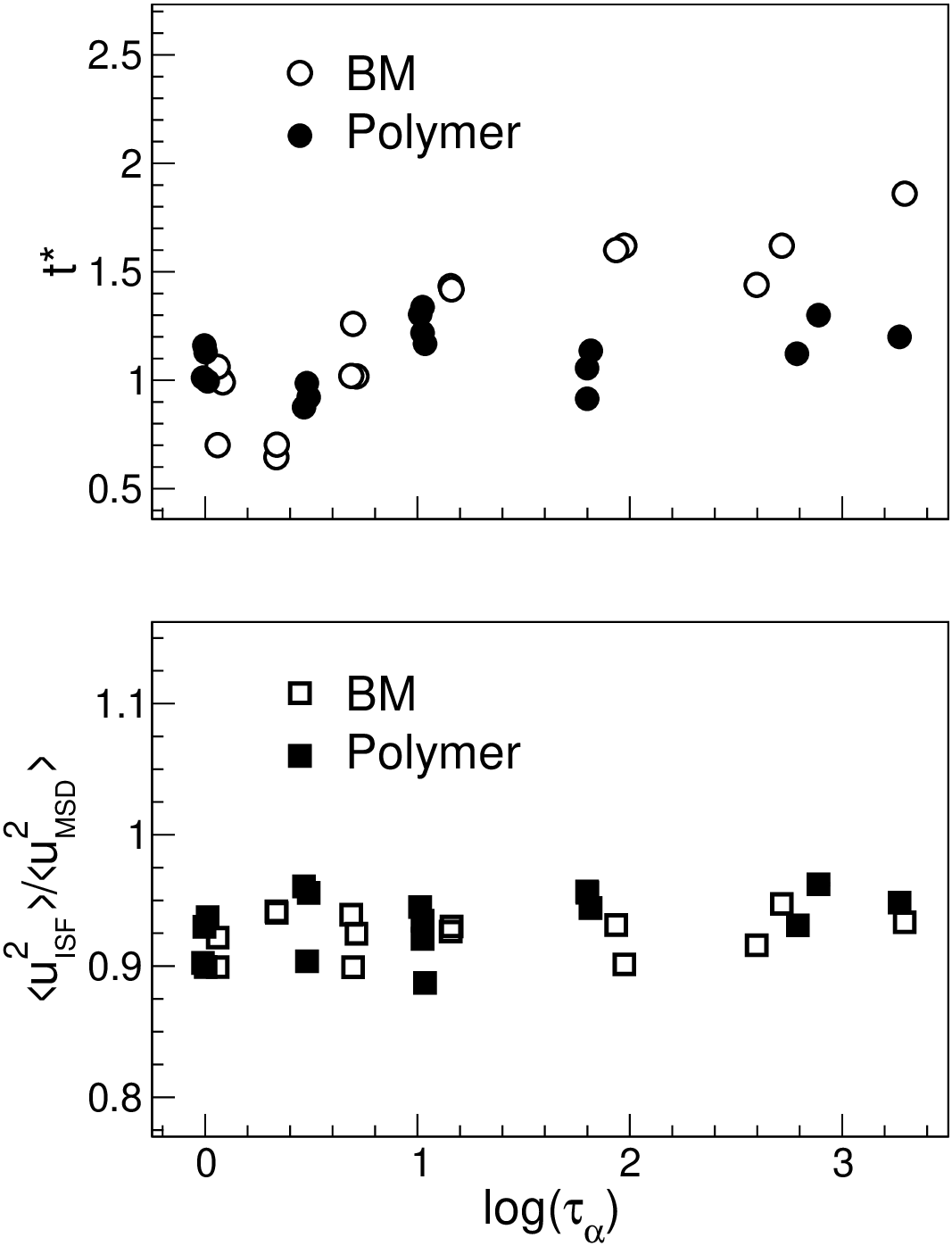}
\end{center}
\caption{Top: position $t^{\star}$ of the minimum slope of MSD in log-log plot. Notice that for polymers $t^{\star}$ is nearly constant , whereas for BM it increases with the structural relaxation time. Bottom: ratio of the DW factor as taken by ISF (Eq.\ref{DWdefISF}) and MSD (Eq.\ref{DWdef}).}
\label{fig6}
\end{figure}

The DW factor is a characteristic length scale of the rattling motion into the cage. The measure of the DW factor must take place in a time window where both the inertial and the relaxation effects are not present. To clearly identify that time window we consider the slope of MSD in the log-log plot
\begin{equation}
\Delta(t) \equiv \frac{\partial \log \langle r^2(t)\rangle}{\partial \log t}
\label{delta}
\end{equation}
Representative plots of $\Delta(t)$ for the polymer system are given in the top inset of Fig.\ref{fig1} and Fig.\ref{fig4}.
$\Delta(t)$ exhibits a clear minimum at $t^{\star} = 1.0(4)$ (corresponding to an inflection point in the log-log plot of $\langle r^2(t)\rangle$) that separates two regimes. The short- and the long-time limits of $\Delta(t)$ correspond to the ballistic ($\Delta(0) = 2$) and the diffusive regimes ($\Delta(\infty) = 1$), respectively. At short times, $t \lesssim 0.7 <  t^{\star}$ the inertial effects become apparent. At long times ($t > \tau_{\alpha} > t^{\star}$) relaxation sets in.  It may be shown that  a minimum of $\Delta(t)$ implies that VCF exhibits a negative tail at long times. A monotonically decreasing VCF, i.e. with no cage effect, leads to a monotonically decreasing $\Delta(t)$. Therefore, MSD at $t^{\star}$ is a mean localization length and the DW factor is defined as 
\begin{equation}
\langle u^2\rangle \equiv  \langle r^2(t=t^{\star})\rangle
\label{DWdef}
\end{equation}
Notice  that  $t^{\star}$, corresponding to about $1-10 ps$ \cite{Kroger04}, is consistent with the time scales of the experimental measurement of the DW factor, e.g. see \cite{BuchZorn92}.
As far as BMs are concerned, the top inset of Fig.\ref{fig4} shows that the time dependence of $\Delta(t)$ is rather similar to the polymer case. Fig.\ref{fig6} (top) plots the position of the minimum of $\Delta(t)$, $t^{\star}$, for the polymer and BM systems. The plot shows that  it is virtually {\it constant} in  polymers, whereas it {\it increases} with the structural relaxation time in BM (see also Fig.\ref{fig4} top panel).

The DW factor is usually experimentally measured by using ISF and considering the height $h$ of the plateau signalling the cage effects (see Fig.\ref{fig1} and Fig.\ref{fig2}) via the relation: 
\begin{equation}
\langle u^2_{ISF}\rangle \; = - \frac{6}{q_m^2} \ln h 
\label{DWdefISF}
\end{equation}
where $h$ is the ISF height at the inflection point of the plateau. Fig.\ref{fig6}(bottom) shows that $\langle u^2\rangle_{ISF}$ and $\langle u^2\rangle_{MSD} \equiv \langle u^2\rangle$ are quite close to each other with  constant ratio within our accuracy. This will have important consequences when comparing the MD simulations with the experimental results in Sec.\ref{scalingexp}. 
\begin{figure}[t]
\begin{center}
\includegraphics[width=8cm]{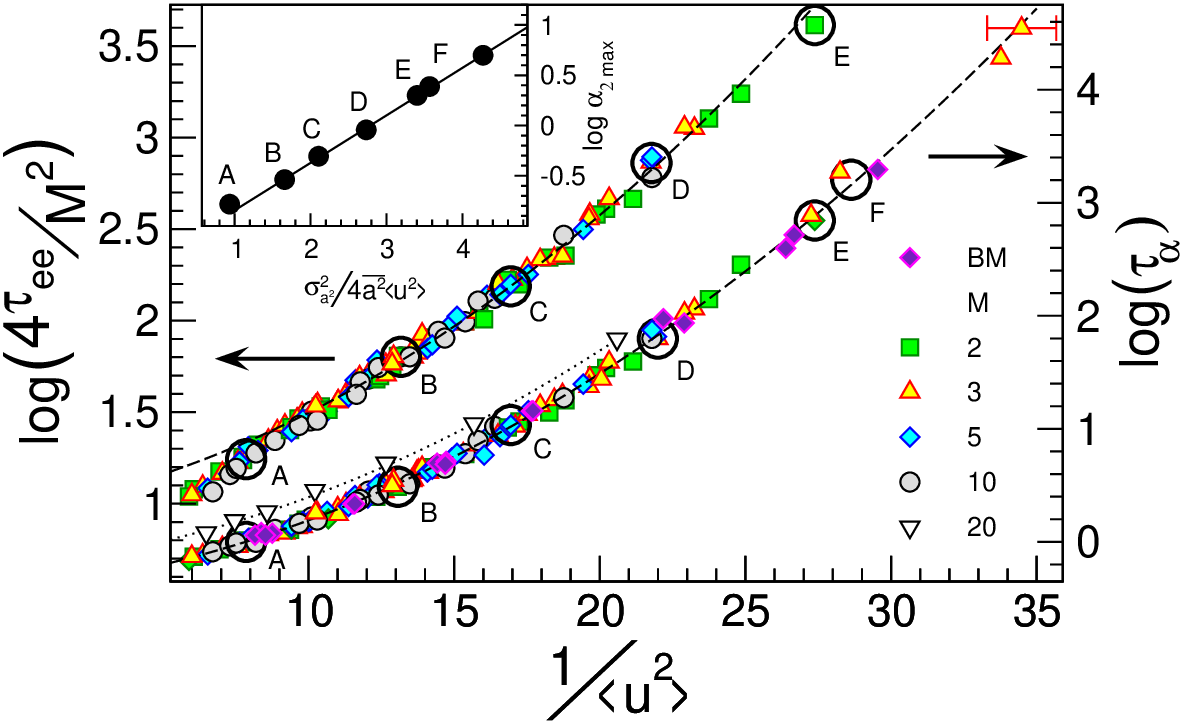}
\caption{The structural relaxation time $\tau_\alpha$ of polymers and BM and the scaled average polymer reorientation time $\tau_{ee}$ vs the DW factor $\langle u^2\rangle$. Empty circles highlight the cases plotted in Fig.\ref{fig1}. For clarity sake the BM cases plotted in Fig.\ref{fig2} are not pointed up. The dashed line across the $\tau_\alpha$ curve is Eq. \ref{parabola} $\log \tau_\alpha = \alpha + \beta \langle u^2\rangle^{-1} + \gamma \langle u^2\rangle^{-2}$ with $\alpha =-0.424(1), \beta = \overline{a^2}/(2 \ln10) = 2.7(1) \cdot 10^{-2}, \gamma = \sigma^2_{a^2}/(8 \ln10) =  3.41(3) \cdot 10^{-3}$. Additional data on the collective relaxation time  $\tau$ are also plotted  ($\triangledown$) \cite{StarrEtAl02}. The dotted curve is obtained by vertically shifting the dashed curve  
($\alpha' = \alpha + 0.205(5)$). The dashed curve  across the scaled average chain reorientation time  curve is a guide for the eyes. Inset: the maximum of the non-gaussian parameter $\alpha_{2 \; max}$ of the A-F clusters of polymer states vs. the ratio of the quadratic and the linear terms of Eq.\ref{parabola} with respect to $\langle u^2\rangle^{-1}$.
}
 \label{fig7}
 \end{center}
\end{figure}

To make it explicit the correlation between the relaxation and the caged dynamics, Fig.\ref{fig7} shows the dependence of both $\tau_\alpha$ and the scaled average chain reorientation time $\tau_{ee}$ on the DW factor. 
The data collapse on two well-defined master curves. The one concerning $\tau_\alpha$ is well fitted by Eq.\ref{parabola}. The master curve of the scaled $\tau_{ee}$ is different. At large DW factor, i.e. fast relaxation, the ratio  $4 \tau_{ee}/ (M^2 \tau_\alpha)$ is roughly constant and {\it decreases}  when the relaxation slow down, as previously reported \cite{BenneEtAl99}. The behaviour at large DW factor is consistent with the Rouse theory concerning the dynamics of unentangled polymers stating that the different relaxation time scales are proportional to each other \cite{DoiEdwards}. The Rouse theory also predicts the scaling $\tau_{ee} \propto M^2$ which is indeed observed even for states with very slowed-down dynamics, see Fig.\ref{fig3}.  However, the differences between the master curves for the chain reorientation and the structural relaxation, which become more apparent for states with sluggish relaxation, evidence one basic limit of that theory, i.e. the assumed gaussian and homogeneous character of the monomer displacements. One anticipates that for slowed down states, where the non-gaussian deviations are large (see Fig.\ref{fig2}) and dynamic heterogeneities are present, the single-monomer relaxation time $ \tau_\alpha$ and the collective relaxation time over a region with size $\sim R_{ee}$, $\tau_{ee}$, cannot be obviously related to each other. We will not analyse further the rotational chain dynamics and henceforth we will focus on the structural relaxation. 

States with different density, chain length and interaction potential are included in Fig.\ref{fig7} corresponding to different degrees of anharmonicity, i.e. non-linear temperature dependence of the DW factor, and then to different fragilities \cite{BordatNgai04,Dyre06,SokolPRL,Angell95,Angell95B,Ngai00,NoviSoko04,NovikovEtAl05}.
The scaling of the structural relaxation time in terms of Eq.\ref{parabola} shows that both the average value $\overline{a^2}$ and the spread $\sigma_{a^2}$ of the square displacements needed to overcome the energy barriers are not affected by the anharmonicity. These parameters  are also not affected by the connectivity, since the master curve collapse data of polymers with different chain lengths and BM.
The best-fit value of the average is $\overline{a^2}^{1/2} \cong  0.35$, consistent with both the observation that $\langle r^2(t=\tau_{\alpha})\rangle^{1/2} \lesssim 0.5$ (see Fig.\ref{fig1}) and the well-known result that the atomic MSD during the structural relaxation is less than one atomic radius ($\sim 0.5$ in MD units) \cite{Angell91}. 

{ The concavity of the master curve in Fig.\ref{fig2} is due to $\sigma_{a^2} \sim 0.25 \neq 0$ indicating the distribution of the displacement required to overcome the energy barriers. Let us show that the concavity is a signature of the heterogeneity of the structural relaxation. In fact, the magnitude of the ratio of the quadratic and the linear terms of Eq.\ref{parabola} with respect to $\langle u^2\rangle^{-1}$,  $\mathcal{R} \equiv \sigma_{a^2}^2/4\overline{a^2}\langle u^2\rangle$, discriminates two different regimes.  If $\mathcal{R} <1$ (large DW), the quadratic term is negligible and  the displacement distribution is not observed being replaced by an effective step length $\overline{a^2}^{1/2}$, i.e. the dynamics is homogeneous. If $\mathcal{R} > 1$ (small DW), the displacement distribution shows up and a heterogeneous mobility distribution is anticipated. 
Indeed, on approaching the glass transition, a spatial distribution of mobilities develops with increasing non-gaussian features \cite{DebeStilli2001,Ngai00,VogGlotz04}, being characterized by the maximum  $\alpha_{2 \; max}$ of NGP \cite{VogGlotz04}. 
For the polymer states in Fig.\ref{fig1}, with NGP shown in Fig.\ref{fig2},  the relation between $\alpha_{2 \; max}$ and $\mathcal{R}$ is shown in the inset of Fig.\ref{fig7}.
It is seen that, when $\mathcal{R}$ exceeds the unit value, $\alpha_{2 \; max}$ increases exponentially. The same is observed for the BM states (not shown).
Notably, the inset of Fig.\ref{fig7} reduces to an activated law for strong glassformers where $\langle u^2\rangle$ is nearly proportional to $T$; this law has been observed for silica \cite{VogGlotz04}.

\subsubsection{Other systems}
\label{scalingother}

\noindent
Eq.\ref{parabola} with the best-fit parameters from  Fig.\ref{fig7} offers the opportunity to find the DW factor $\langle u^2_g\rangle$ at the glass transition of the model polymer and BM system.
 At the glass transition $\tau_{\alpha} = \tau_{\alpha \, g} \equiv 10^2 s$ in laboratory units \cite{Angell91} which corresponds to $\tau_{\alpha \, g}  = 10^{13}-10^{14}$ in dimensionless MD units (the time unit corresponds to $1-10 ps$ \cite{Kroger04}).  Eq.\ref{parabola} yields $\langle u^2_g\rangle^{1/2} = 0.129(1)$. This estimate compares well with other related ones. First, let us consider the ratio between the volume that is accessible to the monomer center-of-mass and the monomer volume is $v_0 \sim (2 \langle u^2_g\rangle^{1/2})^3$. One finds:
 \begin{equation}
v_0 \sim 0.017
\label{freevol}
\end{equation}
\noindent
Flory and coworkers proposed that the glass transition takes place under iso-free volume conditions with the universal value $v_0 \sim 0.025$ \cite{Gedde95}. Furthermore, an extension of the ACS model (leading to the HW equation) predicts that, just as for a crystalline solid \cite{Lowen94}  , there is a Lindemann criterion for the stability of glasses, namely  the ratio 
$ f=\langle u^2_g\rangle^{1/2}/d$, where $d$ is the average next neighbor distance of the atoms in the lattice, is a quasi-universal number ($f  \cong 0.1$) \cite{XiaWolynes00}. Our MD data yield 
\begin{equation}
f^{(MD)} \sim 0.12-0.13
\label{lindenoi}
\end{equation}
where $d$ is taken from the monomer radial distribution function. $f^{(MD)}$ is close to $f= 0.129$ for the melting of a hard sphere fcc solid \cite{Lowen94}. 

\begin{figure}[t]
\begin{center}
 \includegraphics[width=8cm]{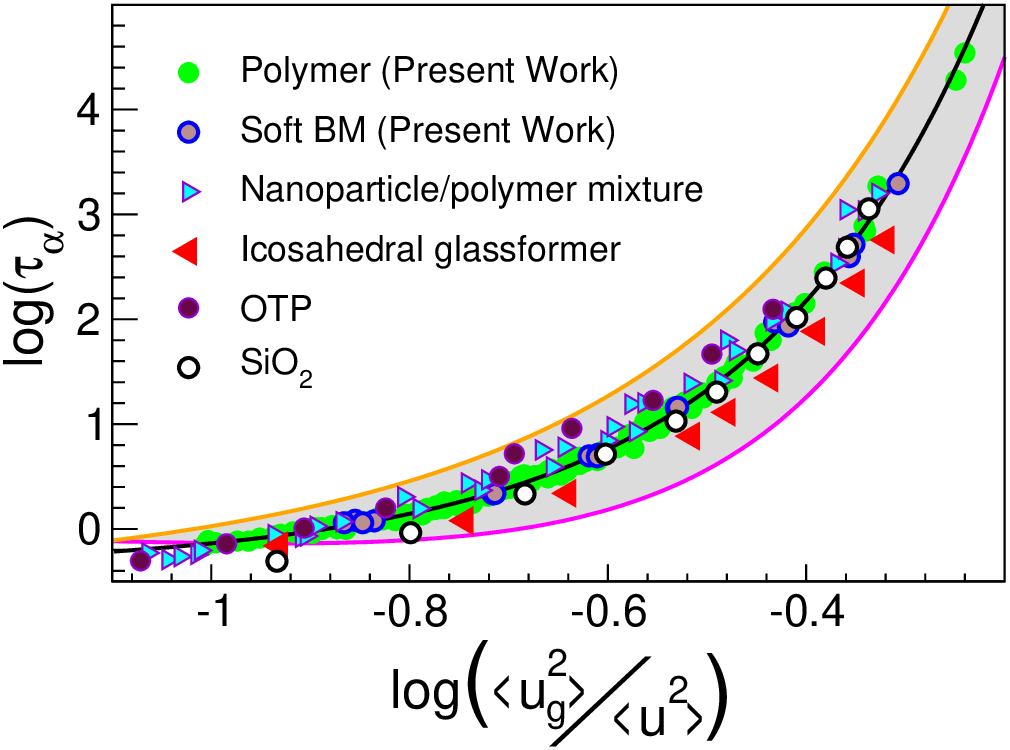}
\end{center}
\caption{Scaling of the structural relaxation time  vs the reduced DW factor of polymers , BM (present work) and other glassformers . The continuous black line is Eq.\ref{scaledparabola}. The colored curves bound  the accuracy of  Eq.\ref{scaledparabola} and correspond to the two definitions $\langle u^2\rangle \equiv  \langle r^2(t=0.6)\rangle$ (magenta, $\langle u^2_g\rangle^{1/2}=0.134(1)$) and $\langle u^2\rangle \equiv  \langle r^2(t=1.4)\rangle$ (orange, $\langle u^2_g\rangle^{1/2}=0.122(1)$). See text for details. 
Data taken from ref.\cite{MSD_SiO}, \cite{ISF_SiO} (Si O${}_2$),   ref. \cite{IcosahedralGlotzer}(icosahedral glassformer),   ref. \cite{NanoparticleDouglas} (Nanoparticle/polymer mixture),  ref.  \cite{OTPRuocco} (OTP, data rescaled to the MD units of the present study). }
\label{fig8}
\end{figure}

The knowledge of $\langle u^2_g\rangle$ allows one to cast Eq.\ref{parabola} in the reduced form:
 \begin{equation}
\log \tau_\alpha = \alpha + \tilde{\beta} \;  \frac{\langle u^2_g \rangle}{\langle u^2 \rangle} + \tilde{\gamma} \left (\frac{\langle u^2_g \rangle}{\langle u^2\rangle} \right )^2 
\label{scaledparabola}
\end{equation}
and using $\alpha,\beta,\gamma$ from Fig.\ref{fig7} yields:  
\begin{eqnarray}
\alpha &=&-0.424(1) \\
\tilde{\beta} &=& \frac{\overline{a^2}}{2 \ln10 \langle u^2_g \rangle}= 1.62(6) \label{betatilde}  \\
\tilde{\gamma} &=&\frac{\sigma^2_{a^2}}{8 \ln10 \langle u^2_g \rangle^2} = 12.3(1) \label{gammatilde}
\end{eqnarray}
The above $\alpha,\tilde{\beta}, \tilde{\gamma}$ values much relies - being the largest MD dataset - on the evaluation of the DW factor of polymers  by setting $t^\star=1$ in Eq.\ref{DWdef}. The uncertainty on  $t^\star$ ($\pm 0.4$, see Fig.\ref{fig1}) leads to an error on $\langle u^2_g\rangle$ and then on $\tilde{\beta}$ and $\tilde{\gamma}$.  One finds $\langle u^2_g\rangle^{1/2}=0.134(1)$ and $\langle u^2_g\rangle^{1/2}=0.122(1)$ for the two extremes $\langle u^2\rangle \equiv  \langle r^2(t=0.6)\rangle$  and $\langle u^2\rangle \equiv  \langle r^2(t=1.4)\rangle$ , respectively. By replacing in Eqs. \ref{betatilde},\ref{gammatilde}  the extremes values of  $\langle u^2_g\rangle$, the bounds setting the accuracy of Eq.\ref{scaledparabola} are found.

Fig.\ref{fig8} compares Eq.\ref{scaledparabola} to the results concerning the polymer and BM systems as well as other model glassformers. $\langle u^2_g\rangle$ values of the latter were evaluated by the extrapolation technique described above, i.e. fitting the raw data by Eq.\ref{parabola} and extrapolating the curve to  $\tau_{\alpha \, g}  = 10^{13}-10^{14}$ in MD units.  It is apparent that, within the accuracy, the scaling procedure works well also in these model glassformers. 

\begin{figure}[t]
\begin{center}
\includegraphics[width=8cm]{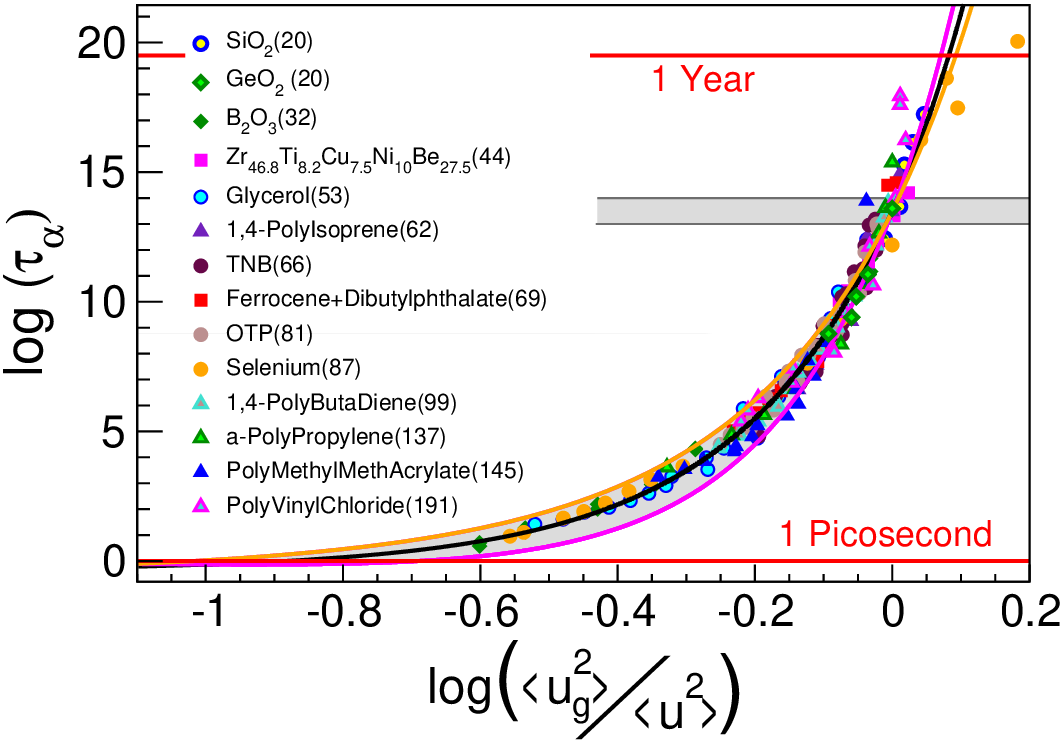}
\end{center}
\caption{Scaling of the structural relaxation time $\tau_\alpha$ (in MD units) vs. the reduced DW factor by considering the experimental results  concerning several glassformers. The grey area marks the glass transition. The continuous black line is the scaling curve Eq.\ref{scaledparabola}. The numbers in parenthesis denote the fragility $m$. The error bounds on the scaling curve are the same of Fig.\ref{fig8}.  Data sources are listed in ref.\cite{OurNatPhys}. Ge O${}_2$ data from ref.\cite{GeO2Fontana}.}
\label{fig9}
\end{figure}

\subsubsection{Experiments}
\label{scalingexp}

Eq.\ref{scaledparabola} is well-suited for comparison with  the available experimental data. It is important to note that, even if Eq.\ref{scaledparabola} is derived in terms of $\langle u^2\rangle_{MSD} \equiv \langle u^2\rangle$, in view of the constant ratio $\langle u^2\rangle_{ISF}/\langle u^2\rangle_{MSD}$ (see Fig.\ref{fig6}), it also holds for $\langle u^2\rangle_{ISF}$, the quantity which is usually provided by the experiments. 

Fig.\ref{fig9} compares the master curve Eq.\ref{scaledparabola} with the experimental data on several glassformers and polymers in a wide range of fragility. It covers a range of relaxation times from picoseconds to almost one year}. The scaling in Fig.\ref{fig9} cannot be ascribed to $\langle  u^2_g \rangle$ which weakly correlate with the fragility $m$, see Fig.\ref{fig10}. Instead, it shows that both the reduced mean square displacement $\overline{a^2}/\langle  u^2_g \rangle$ to overcome the energy barriers and the related spread  $\sigma_{a^2}/\langle  u^2_g \rangle$ are fragility-independent, and then also the curvature of the master curve.

The experimental data in Fig. \ref{fig3} were collected by changing the temperature. In this respect, the universal scaling of Fig.\ref{fig3} proves that the well-known increasing deviation of $\langle  u^2(T)\rangle$ from the linear temperature dependence of the harmonic behavior by increasing the fragility index $m$ \cite{Angell95,Ngai00,Ngai04,NoviSoko04,NovikovEtAl05} just mirrors the corresponding increasing bending of $ \tau_\alpha(T)$ vs $T_g/T$ in the Angell plot \cite{Angell91} from the glass transition region up to the liquid state. However, the glass transition may be reached under isothermal conditions also by increasing  the density or the connectivity (here expressed by the chain length) \cite{PrevostoEtAl04}. Our MD results highlights the correlation of structural relaxation and vibrational dynamics also for these alternative routes which awaits experimental confirmation.
\begin{figure}[t]
\begin{center}
\includegraphics[width=8cm]{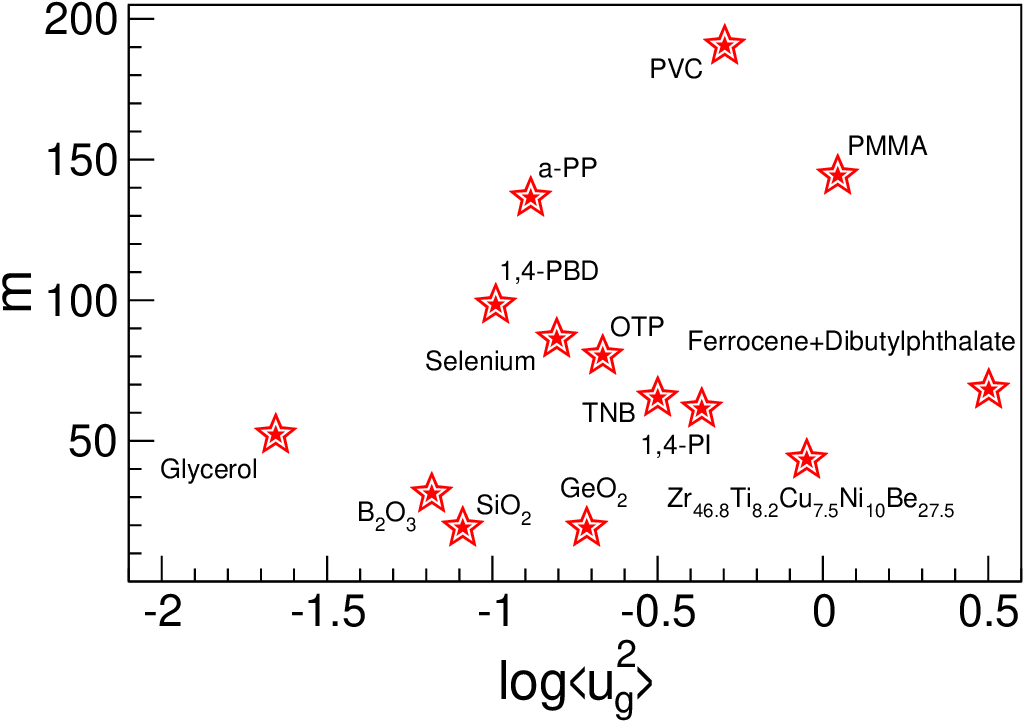}
\end{center}
\caption{Correlation plot between fragility and DW factor at the glass transition.}
\label{fig10}
\end{figure}

\subsubsection{Comparison with other scaling procedures, Lindemann criterion}
\label{scalingcomparison}

Several scaling and correlation plots of the structural relaxation of glassforming systems were reported. They belong to two classes: one class, as the present approach, considers the relaxation times (or viscosity) of states close to and {\it far from} the glass transition \cite{kivelson,novikovEL96,RosslerEtAlJNCS98,blochowicz06}. We will refer to that as {\it $\tau_\alpha$-plots}. The other class considers the fragility, i.e. the behaviour {\it close} to the glass transition \cite{fujimori95,ScopignoEtAl03,NoviSoko04,NovikovEtAl05,Buchenau04,MarAngell01}. These will be referred to as  {\it fragility-plots}. Customarily, the alternative scaling procedures considered only data where the approach to the glass transition occurs by changing the temperature at ambient pressure, whereas the robustness of the present DW scaling to pressure changes was validated by MD simulations. 
\cite{RosslerEtAlJNCS98}
\begin{table}[t]
\begin{center}
  \begin{tabular}{|c|c|c|c|c|c|}
    \hline 
   \multirow{2}{*}{Ref.}  & \multirow{2}{*}{$\tau_\alpha^{min}$ (ns)} & \multirow{2}{*}{$m_{min}$}  & \multirow{2}{*}{$m_{max}$} & \multirow{2}{*}{Polymers} & Adjustable  \\
    & & & &  & Parameters\\
    \hline
    PW & $\sim 0.001$ & $20$ & $191$ & Y ($6$)  & $1$ \\ 
    \hline
    \cite{kivelson} & $\sim 0.001$ & $32$ & $160$ & Y ($3$)  & $3$ \\
    \hline
    \cite{novikovEL96,RosslerEtAlJNCS98} & $\sim 1$ & $25$ & $102$ & N  & $1$\\
    \hline
    \cite{novikovEL96,RosslerEtAlJNCS98} & $\sim 1000$ & $71$ & $174$ & Y ($8$)  & $1$ \\
    \hline 
    \cite{blochowicz06} & $ \sim 10$ & $53$ & $124$ & N  & $1$ \\
    \hline   
    \end{tabular}
\end{center}
\caption{Comparison of the $\tau_\alpha$-scaling procedures in literature with the present work (PW).
$m_{min}$ and $m_{max}$ are the minimum and the maximum fragility
of the glass formers considered. For $\tau_\alpha < \tau_\alpha^{min}$ the scaling fails. All the scaling procedures include the glass transition region.}
\label{taualfaplots}
\end{table}

A comparison between the present analysis and other {\it $\tau_\alpha$-plots} is presented in Table \ref{taualfaplots} which lists the number of adjustable parameters to build up the master curve, the shortest relaxation time $\tau_\alpha^{min}$ below which the scaling fails, the fragility range being covered and the possible inclusion of polymers. It must be pointed out that the DW scaling adjusts only the conversion factor between the MD and the actual time units, i.e. the vertical shift factor. Within the errors, this factor is nearly {\it independent} of the system, with the notable exception of $B_2 O_3$ (2d-sheet structure) \cite{OurNatPhys}.

Fragility plots assume that fragility, i.e. the slope of the curve $\log \tau_\alpha$ vs. $T_g/T$ at $T_g$, is a distinctive characteristic of glass-forming systems. As a consequence, they  involve much less data than $\tau_\alpha$-plots. A comparison between the present scaling and some fragility-plots is presented in Table \ref{fragplots}.
\begin{table}[t]
\begin{center}
  \begin{tabular}{|c|c|c|c|c|c|c|}
    \hline 
   \multirow{2}{*}{Ref.}  & \multicolumn{2}{|c|}{Data} & \multirow{2}{*}{$m_{min}$}  & \multirow{2}{*}{$m_{max}$} & \multirow{2}{*}{Polymers} & Adjustable  \\
    \cline{2-3}
    & Exp & Sim &  &  & &  Parameters\\
    \hline
    PW & $184$ & $120$ & $20$ & $191$ & Y ($6$) & $1$ \\ 
    \hline
   \cite{fujimori95} & $4$ & - & $20$ & $90$ & N & $2$ \\
    \hline
   \cite{ScopignoEtAl03} & $10$ & - & $20$ & $87$ & Y ($2$) & $1$\\
    \hline
     \cite{NoviSoko04} & $15$ & - & $20$ & $100$ & N & $2$ \\
    \hline 
   \cite{MarAngell01} & $24$ & - & $20$ & $160$ & N & $1$ \\
    \hline   
    \end{tabular}
\end{center}
\caption{Comparison of the fragility-scaling procedures in literature with the present work (PW).}
\label{fragplots}
\end{table}

Recently, Niss {\it et al} considered the HW relation, Eq.\ref{dyreWolynes} \cite{NissEtAl09}. By assuming that the intermolecular distance scales with the density as $\rho^{-1/3}$, Eq.\ref{dyreWolynes} was recast as  :
\begin{equation}
\label{dyredensity}
\tau_\alpha^{(N)}, \eta^{(N)} \propto  \exp\left(\frac{C \rho_g^{-2/3}}{\langle u^2 \rangle}\right)
\end{equation}
where $\rho_g$ and $C$ are the density at the glass transition and a constant, respectively. It was concluded that, if one {\it defines} the glass-transition Lindemann ratio as $f^{(N)} \equiv \rho_g^{2/3} \langle u^2_g \rangle$, the latter is system-dependent, i.e. it is not universal. It is interesting to investigate the quantity $f^{(N)}$  for the polymer and BM models under study. It will be shown that, consistently with ref.\cite{NissEtAl09}, $f^{(N)}$ is system-dependent. This suggests that $f^{(N)}$ is a less promising definition of the Lindemann ratio than $f^{(MD)}$ which, according to the present simulation, depends very weakly on the system, see Eq.\ref{lindenoi}. It must be reminded that our MD data set correspond to different kind of polymeric and BM systems in that different interacting potentials, Eq.\ref{Eq:modifiedLJ}, are considered and, for polymers, different chain lengths.
To test the quantity $f^{(N)}$, the correlation plot between $\log \tau_\alpha$ and $ \rho^{-2/3} \langle u^2\rangle^{-1}$ was fitted with:
\begin{equation}
\log {\tau_\alpha}^{(N)} = C_1 + C_2 \rho^{-2/3} \langle u^2\rangle^{-1} + C_3 \rho^{-4/3} \langle u^2\rangle^{-2}
\label{dyredensityquad}
\end{equation}

\begin{figure}[t]
\begin{center}
\includegraphics[width=8cm]{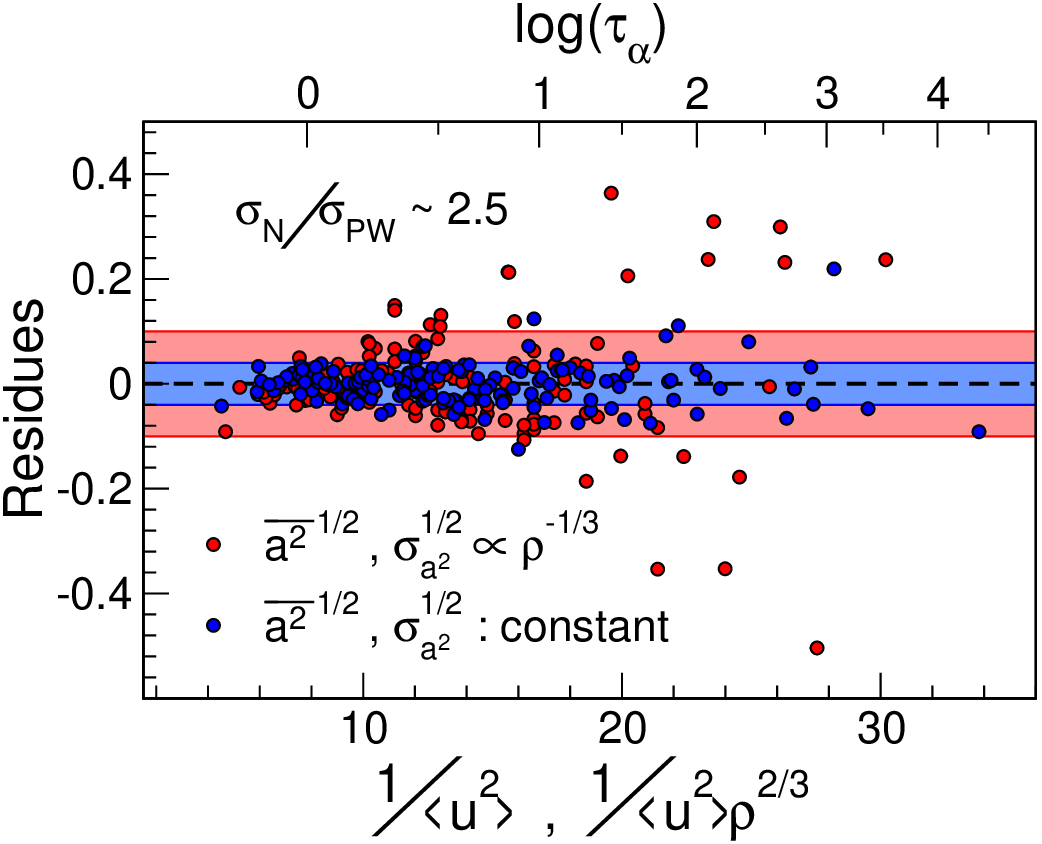}
\end{center}
\caption{Residues of the best-fit of  $\log \tau_\alpha$ vs. $ \langle u^2\rangle^{-1}) $ with Eq.\ref{parabola}  (blue dots) and $\log \tau_\alpha$ vs. vs. $\rho^{-2/3} \langle u^2\rangle^{-1}$ with Eq.\ref{dyredensityquad}  (red dots). The ratio between the two standard deviations $\sigma_N/ \sigma_{PW}$ is indicated. Each colored band spans $\pm \sigma$. }
\label{fig11}
\end{figure}

The above form is analogous to Eq.\ref{parabola} and suitably generalizes Eq.\ref{dyredensity} to account for the bending of the plot of $\log \tau_\alpha$ vs. $\rho^{-2/3} \langle u^2\rangle^{-1}$ due, in turn, to the bending of $\log \tau_\alpha$ vs $\langle u^2 \rangle^{-1}$ (see Fig.\ref{fig7}).   Eq.\ref{dyredensityquad} assumes that the density scaling of the characteristic length scales fulfills  the ansatz $ {\overline{a^2}}^{1/2}, {\sigma_{a^2}}^{1/2} \propto \rho^{-1/3}$. Instead, Eq.\ref{parabola} takes both quantities as {\it constant}.
Fig.\ref{fig11} compares the residues of the fit
of $\log \tau_\alpha$ vs $\langle u^2 \rangle^{-1}$ with Eq.\ref{parabola} (see Fig.\ref{fig7}) and the fit of $\log \tau_\alpha$ vs $ \rho^{-2/3} \langle u^2\rangle^{-1}$ with Eq.\ref{dyredensityquad}. Both fits have the same number of adjustable parameters.  It is seen that the residues are structureless, i.e. the fits have equal accuracy. However, the discrepancies from Eq.\ref{dyredensityquad} exhibit standard deviation $\sigma_N$ which is larger than the one, $\sigma_{PW}$, of the deviations from Eq.\ref{parabola}. Note also that the deviations from Eq.\ref{dyredensityquad} increase with $\tau_\alpha$. 
When an extrapolation procedure analogous to the one outlined in Sec.\ref{scalingother} is followed to derive $\rho^{2/3} \langle u^2_g\rangle$, i.e. $f^{(N)}$, the poorer collapse of the data when $\log \tau_\alpha$ is plotted vs. the quantity $ \rho^{-2/3} \langle u^2\rangle^{-1}$ results in a $f^{(N)}$ value with larger uncertainty, i.e. less "universal", than $f^{(MD)}$.  

\section{Conclusions}
\label{Conclusions}
The paper presents a thorough analysis of the scaling between the long-time relaxation and the caged dynamics. MD simulations of both a model polymer system and a binary mixtures were carried out by varying the temperature, the density, the potential and the polymer length to consider the structural relaxation as well as both the rotational and the translation diffusion. They showed the existence of different physical states exhibiting coinciding transport and relaxation from the end of the ballistic regime through the diffusive one. This points to a link between the short- and the long-time dynamics which is evidenced by the master curves found by
correlating  the DW factor with the structural relaxation  time $\tau_{\alpha}$ and the chain rotational diffusion.
An analytic model of $\tau_{\alpha}$ master curve is developed, leading to Eq.\ref{parabola}, which fits nicely with the MD results on polymers and BM. Notably, the model does not predict the existence of physical states yielding the divergence of $\tau_\alpha$.
By using suitable reduced units, the MD $\tau_{\alpha}$ scaling on polymers and BM was extended  to include MD data on other systems as well as the experimental data on several glassformers and polymers in a wide range of fragility by covering a range of relaxation times from picoseconds to several days. The scaling in terms of the DW factor compare favourably with other scaling procedures. In particular, the density scaling of the characteristic length scales according to the ansatz $ {\overline{a^2}}^{1/2}, {\sigma_{a^2}}^{1/2} \propto \rho^{-1/3}$ is not supported by the present simulations. The study suggests that the equilibrium and the moderately supercooled states of the glassformers possess key information on the huge slowing-down of their relaxation close to the glass transition which, according to our simulations, exhibits features shared with the Lindemann melting criterion and the free-volume model.

\acknowledgments
S.Capaccioli, J.F.Douglas, G. P. Johari and M.Malvaldi are warmly thanked for discussions. Financial support from MUR within the PRIN project ``Aging, fluctuation and response in out-of-equilibrium glassy systems'' and FIRB project ``Nanopack'' as well as computational resources by ``Laboratorio per il Calcolo Scientifico'', Pisa are gratefully acknowledged.

\bibliography{wolynjp}
\end{document}